\documentstyle[aps,preprint,epsfig]{revtex}

\begin{document}
\draft
\title{Bursts in the Chaotic Trajectory Lifetimes 
Preceding the Controlled Periodic Motion}

\author{V. Paar and H. Buljan}
\address{Department of Physics, Faculty of Science, University of
Zagreb, 10000 Zagreb, Croatia}
\date{\today}

\maketitle

\begin{abstract}
The average lifetime ($\tau(H)$) it takes for a randomly started 
trajectory to land in a small region ($H$) on a chaotic 
attractor is studied. 
$\tau(H)$ is an important issue for controlling chaos. 
We point out that if the region $H$ is visited by a short periodic 
orbit, the lifetime $\tau(H)$ strongly 
deviates from the inverse of the naturally invariant measure 
contained within that region ($\mu_N(H)^{-1}$). 
We introduce the formula that relates $\tau(H)/\mu_N(H)^{-1}$ 
to the expanding eigenvalue of the short periodic orbit 
visiting $H$. 
\end{abstract}

\pacs{05.45.Gg, 05.45.Ac, 05.45.-a}

Controlling chaos by stabilizing one of the many unstable 
periodic orbits embedded within a given chaotic 
attractor is attainable with small, time-dependent changes 
in an accessible system parameter\cite{OGY_1,DRS,O_b}. 
The idea is to observe a typical trajectory of the 
uncontrolled system for some transient time, 
until it falls sufficiently close to the desired 
periodic orbit, and then to activate the control 
mechanism. 
An important issue related to the utilization of 
this method is the average lifetime of chaotic transients 
that precede the controlled periodic motion
\cite{OGY_1,O_b,Shin_1,Shin_2}.

Suppose that the uncontrolled chaotic attractor $A$ 
describes the asymptotic behavior of the dynamical 
system $O:D\rightarrow D,\ D\subseteq 
\mbox{$\mathbf{R}$}^m$, also referred to as the 
original system. 
Let $\vec\xi=O^{k}(\vec\xi)$ be a point 
on a particular unstable periodic orbit\cite{Aue,Eck} 
that we wish to stabilize. 
Furthermore, let the vicinity of the orbit 
$H\equiv H_{\epsilon}(\vec\xi)$ be an $m$-dimensional 
ball of radius $\epsilon<<1$ centered at $\vec\xi$. 
The probability that a randomly 
started trajectory does not reach $H$ up to 
time $t$ is $\sim e^{-t/\tau(H)}$. 
The average lifetime $\tau\equiv\tau(H)$ is strongly 
correlated with the visitation frequency of typical trajectories 
to the region $H$, which is described in terms of 
the naturally invariant measure ($\mu_N$) contained 
within $H$ - $\mu_N(H)$\cite{OGY_1,Shin_1,GORY}. 
Obviously, if a certain region on a given chaotic attractor 
is visited more frequently by typical trajectories, 
the average lifetime it takes for an orbit to land in 
that region will be smaller. 
In the present study we address the following question: What 
is the deviation of $\tau$ from $\mu_N(H)^{-1}$ as 
a function of $\vec \xi$ and $\epsilon$?

We will demonstrate the existence of bursts in the lifetimes, 
i.e. significant deviations of $\tau$ from $\mu_N(H)^{-1}$, which 
appear when the $H$ region encompasses a point 
on a short periodic orbit. 
In contrast to the overall $\tau\simeq\mu_N(H)^{-1}$ 
behavior, at these exceptional positions, the lifetime 
$\tau$ is considerably prolonged as compared to 
$\mu_N(H)^{-1}$. 
As the length of the shortest cycle 
visiting $H$ increases, 
the parameter of this deviation, $S(H)\equiv\tau/\mu_N(H)^{-1}$, 
decreases rapidly towards $1$. 
We will introduce a formula that relates the parameter $S(H)$
to the repelling properties (expanding eigenvalue) of the 
shortest cycle within $H$. 
Furthermore, we will demonstrate that $S(H)$ is 
independent of $\epsilon$ (for $\epsilon <<1$). 
This is consistent with the previously 
reported scaling 
$\tau\sim\mu_N(H)^{-1}$(see e.g. \cite{OGY_1,Shin_1}).

The present study is motivated by the previous investigation 
of the logistic map with a hole\cite{PP1}. 
In this paper we present a theoretical explanation for 
the phenomenological result reported in Ref. 
\cite{PP1} and generalize it to 1D noninvertible and 
2D invertible chaotic maps. 
(From our considerations a conjecture follows that 
similar phenomena occur generally in chaotic systems.)

It will be useful to define an 
auxiliary modified map \cite{PP1,PP2}

\begin{equation}
M(\vec\xi')=\left\{
         \begin{array}{ll}
          O(\vec\xi') , & \mbox{$\vec\xi'\in D\backslash H$} \\ 
	  \mbox{outside of the basin of } A , & \mbox{$\vec\xi'\in H$}.
         \end{array}
        \right.    \label{M}
\end{equation}
A typical trajectory of the map $O$ remains on the chaotic attractor 
forever,  while the same trajectory in the map $M$ 
eventually escapes through the region $H$, from now on 
also referred to as the hole. 
The average lifetime of chaotic transients 
created by the map $M$ is equal to 
$\tau(H)$, which we have defined above. 
Similar maps with a forbidden gap region arise in the context 
of communicating with chaos\cite{JOH}, and in calculation of the 
topological entropy\cite{Cv}.

To illustrate the concept of bursts, 
we consider two chaotic 1D maps: 
(i) the asymmetric tent map 
$O(x)=k_1 x,\ x<k_1;\ O(x)=k_2 (1-x),\ x>k_1$, 
$k_1,k_2>0$, $k_1^{-1}+k_2^{-1}=1$, 
and 
(ii) the sinusoidal map $O(x)=\sin{\pi x}$. 
Fig. \ref{brs_ill} displays $\tau$ as a function of 
the position of the hole $\xi$ ($H=(\xi-\epsilon,\xi+\epsilon)$), 
for the two paradigmatic maps (see also Ref. \cite{PP1}). 
The width of the hole is kept constant $(\epsilon=0.005)$.
Both graphs exhibit some common features: 
(i) the overall behavior of lifetimes follows the $\mu_N(H)^{-1}$ 
pattern; 
(ii) strong local deviations from the $\mu_N(H)^{-1}$ behavior - 
the bursts, observed as leaps in the 
lifetimes, occur when the hole interval sweeps across 
a short periodic orbit; 
(iii) the bursts are more significant if the length 
of the short periodic orbit is smaller.

The explanation of the burst phenomenon requires the 
comparison of two concepts: 
(i) the conditionally invariant measure \cite{Pian,Tel,Telc} 
(also referred to as the $c$-measure) - the concept associated 
with the modified system, and 
(ii) the naturally invariant measure \cite{Eck,O_b} 
of the original system. 
In order to define these measures, 
imagine that we cover the chaotic attractor with cells $(I)$ 
from a very fine grid. 
Then we randomly distribute a large number $(N)$ of 
points on the grid, and evolve them under the dynamics $O$ for a 
long time $T$. 
Suppose that all initial points are colored blue, 
and that a point irretrievably 
changes color from blue to red immediately after its 
first entrance into the region $H$. 
Thus, the point $\vec x_T=O^T(\vec x_0)$ at time $T$ 
is blue if $O^t(\vec x_0)\notin H$ for $t\in \{ 0,1,\ldots,T-1\}$, 
and red otherwise. 
In the limit $T\rightarrow\infty$, the fraction of points  
found in a given cell $I$, is just the natural measure 
contained within that cell: 
$\mu_N(I)=N_T(I)/N;\ N=\sum_I N_T(I)$, 
where $N_T(I)=b_T(I)+r_T(I)$ denotes the total 
number of points in a given cell $I$ at time $T$
\cite{O_b}. 
The number of blue (red) points within a given 
cell $I$ at time $T$ is denoted as $b_T(I)\ (r_T(I))$. 
The points which change color from blue to red under 
the action of the map $O$, are those which would escape 
the attractor under the action of the map $M$. 
Hence, if we were to evolve exactly the same initial conditions 
using the map $M$ for the same time $T$, 
the number of surviving points (blue points) would be 
$B_T=\sum_I b_T(I)\sim N\exp{(-T/\tau)}$. 
In the limit $N\rightarrow\infty$, $T\rightarrow\infty$, 
the distribution of blue points 
converges to the $c$-measure of the modified system
\cite{Pian,Tel,Telc}. 
The fraction of blue points in a 
given cell $I$ is simply the $c$-measure contained within 
that cell: $\mu_C(I)=b_T(I)/B_T$.

The blue point at any time $t>0$, was certainly not in the hole 
$H$ at time $t-1$. 
Therefore, $M^{-1}(I)\equiv O^{-1}(I)\backslash H$ 
and $b_T(I)=b_{T-1}(M^{-1}(I))$, 
which divided by $B_T=B_{T-1}\exp{(-1/\tau)}$ yields the 
well known relation for the $c$-measure 
\cite{Pian,Tel,Telc}

\begin{equation}
\mu_C(I)=e^{\frac{1}{\tau}} \mu_C(M^{-1}(I)). \label{cinv}
\end{equation}
By summing the equation above over all the cells $I$ we obtain 

\begin{equation}
\frac{1}{\tau}=-\ln [1-\mu_C(H)]
             \simeq \mu_C(H). \label{tauC}
\end{equation}
We emphasize the importance of this observation. 
The average lifetime it takes for a typical trajectory to 
reach the small region $H$ on the attractor is an inverse of 
the $c$-measure contained within that region ($\mu_C(H)^{-1}$), 
which may significantly differ from the inverse of the 
natural measure ($\mu_N(H)^{-1}$).

As an illustration, in Fig. \ref{c_xsin} we display the $c$-measure 
for the modified version of the map $\sin{\pi x}$, 
in comparison to the natural measure
of the original map.
For the $c$-measure in Fig. \ref{c_xsin} a), 
the hole has been positioned at an arbitrary point, 
but not on the short periodic orbit. 
In this case, we observe that $\mu_C(H)\simeq\mu_N(H)$, i.e.,  
$\mu_N(H)^{-1}$ is a good approximation for 
the lifetime. 
In contrast, in Fig. \ref{c_xsin} b) we display $\mu_C$ 
for the modified map $\sin{\pi x}$, with the hole positioned 
on the fixed point. 
We notice that $\mu_C(H)$ strongly deviates 
from $\mu_N(H)$. 
This case corresponds to the burst labeled $1$ in Fig. \ref{brs_ill} b). 
The overall agreement of the two measures is evident 
in both Figures \ref{c_xsin} a) and b). 
However, at locations above the first few images of the hole, 
$\mu_C$ takes the shape of a well, with values which are 
considerably lower than $\mu_N$. 
When the hole lies on the fixed point (Fig. \ref{c_xsin} b)), 
the wells are just above the hole itself. 
This results in a pronounced deviation of $\tau=\mu_C(H)^{-1}$ from 
$\mu_N(H)^{-1}$, which manifests as a burst.

Now we compare the two measures globally. 
A chaotic repeller is a set of points on the 
attractor that never visit the hole\cite{Tel,Telc,JOH}. 
A trajectory that starts close to the repeller, 
does not escape the attractor for a long time. 
Therefore, the blue points at a large time $T$ are 
located along the unstable manifold of the 
repeller. 
Their distribution along this manifold 
defines the $c$-measure\cite{Tel}. 
Thus, the natural measure is constructed from all the points 
(at time $T$) on all the unstable manifolds, 
whereas the $c$-measure results only 
from points on parts of these manifolds. 
The parts which extend from the repeller up to the hole. 
As we reduce the size of the hole $\epsilon$, 
the repeller and its unstable manifold grow. 
Consequently, $\mu_C$ gradually approaches $\mu_N$, 
and for sufficiently small $\epsilon$, the two measures 
are practically identical. 
(For $\epsilon=0$, $\mu_C$ becomes $\mu_N$\cite{Pian,Tel,Telc}.)

However, the deviation of the lifetime $\tau=\mu_C(H)^{-1}$ from 
$\mu_N(H)^{-1}$ depends only on the values 
of the two measures within the hole, and therefore is 
a local quantity. 
In order to make a more accurate comparison
of $\mu_C$ and $\mu_N$, we introduce the following 
definitions. 
Consider a set $P\subset D$ such that 
$\mu_N(P)>0$. 
We define the quantity 

\begin{equation}
a(P)=\mu_C(P)/\mu_N(P), 
\end{equation}
which describes the relation between 
$\mu_C$ and $\mu_N$ within $P$. 
We also define the influence $i(P)$ of the 
hole on the set $P$ as 

\begin{equation}
i(P)=\frac
{\mu_N(O^{-l(P)}(P)\cap H)}{\mu_N(O^{-l(P)}(P))}.
\end{equation}
$l(P)$ denotes the smallest integer for which the 
section $O^{-l(P)}(P)\cap H$ becomes nonempty. 
The natural invariant measure within $O^{-l(P)}(P)$ is 
mapped to $P$ in $l(P)$ iterates. 
The influence is just a fraction ($0 \leq i(P) \leq 1$) 
of $\mu_N(P)$ that is mapped from the hole in the 
last $l(P)$ time steps.

Let $P_{\epsilon}\equiv P_{\epsilon}(\vec x)\subset D$ 
be an $m$-dimensional ball of radius $\epsilon$
(the same radius as the hole) centered 
at $\vec x$. 
We ask the following question: 
Given a chaotic attractor and choosing the hole region, 
what is the behavior of 
$a(P_{\epsilon})=\mu_C(P_{\epsilon})/\mu_N(P_{\epsilon})$ 
as the position of $P_{\epsilon}$ on the attractor 
is changed?

By using Eq. (\ref{cinv}) and the identity 
$\mu_N(P_{\epsilon})=\mu_N(O^{-1}(P_{\epsilon}))$, 
we can write

\begin{equation}
a(P_{\epsilon})=e^{\frac{l}{\tau}}\cdot
(1-i(P_{\epsilon}))\cdot 
a(M^{-l}(P_{\epsilon})), \label{recur} 
\end{equation}
where $l\equiv l(P_{\epsilon})$ 
(in what follows, $l\equiv l(P_{\epsilon})$).

Concerning the first factor in Eq. (\ref{recur}), note that 
the average lifetime typically scales like 
$\tau\sim 1/\epsilon^{D_p(\vec\xi)}$
($D_p(\vec\xi)$ denotes the pointwise dimension at $\vec\xi$)
\cite{OGY_1,Shin_1,O_b}, 
whereas the minimal number of iterates $l$ 
for which $O^{-l}(P_{\epsilon})\cap H\neq \emptyset$ 
scales like $l\sim \ln(1/\epsilon)$\cite{Shin_1}. 
Therefore, $\exp(l/\tau)\simeq 1+l/\tau\simeq 1$.

If the influence $i(P_{\epsilon})$ is 
small, the second factor in Eq. (\ref{recur}) is 
$\simeq 1$.
We argue that $i(P_{\epsilon})$, the influence of the 
hole on the region $P_{\epsilon}$ decreases 
exponentially with $l$. 
For the 2D original map, $O^{-l}(P_{\epsilon})$ 
is a narrow region which is stretched along the stable direction 
and squeezed along the unstable one\cite{Shin_1}. 
The intersection of $O^{-l}(P_{\epsilon})$ with the hole $H$ 
is roughly a rectangle of length $\epsilon$ and 
width $\epsilon\exp(-\lambda_1 l)$. 
For the 1D map, $O^{-l}(P_{\epsilon})\cap H$ 
is an interval of width 
$\sim\epsilon\exp(-\lambda_1 l)$. 
In both cases, $\lambda_1$ denotes the positive Lyapunov 
exponent obtained for typical initial conditions 
on the attractor. 
Since the natural measure is concentrated along 
the unstable manifolds\cite{Eck,O_b}, we can relate 
$\mu_N(O^{-l}(P_{\epsilon})\cap H)\sim 
\exp(-\lambda_1 l)$. 
Thus, due to the chaoticity of the map $O$ 
we obtain $i(P_{\epsilon})\sim \exp(-\lambda_1 l)$.

Concerning the third factor in Eq. (\ref{recur}), 
we consider the set $M^{-l}(P_{\epsilon})$ and the value 
$a(M^{-l}(P_{\epsilon}))$ in dependence of $l$. 
For the 2D maps, the set $M^{-l}(P_{\epsilon})\equiv 
O^{-l}(P_{\epsilon})\backslash H$ is stretched 
exponentially fast with increasing $l$ along the stable 
manifolds, and thus crosses many of the unstable manifolds 
that carry both the natural and the $c$-measure. 
For the 1D maps, the number of disjoint intervals that 
make the $l-th$ preimage of $P_{\epsilon}$ grows 
exponentially with $l$. 
Furthermore, they are scattered all over the 
attractor. 
Due to the chaoticity of the map $O$, 
in both cases 
$M^{-l}(P_{\epsilon})$ becomes more 
democratic with larger $l$, in the sense that the 
value $a(M^{-l}(P_{\epsilon}))$ reflects the global 
agreement between the two measures. 
Thus, insofar as $l$ is not small, 
$a(M^{-l}(P_{\epsilon}))\simeq 1$.

We conclude that for $l\equiv l(P_{\epsilon})$ larger than some 
critical value (call it $l_c$), all of the three factors in 
Eq. (\ref{recur}) are $\simeq 1$ and therefore 
$\mu_C(P_{\epsilon})\simeq\mu_N(P_{\epsilon})$. 
This is consistent with the global agreement between the two measures.
The critical value $l_c$ depends on the chaoticity of the 
original map. 
For example, we may take $l_c$ to be the 
smallest integer for which $e^{-\lambda_1 l_c}<0.1$
(e.g. for the $\sin(\pi x)$ map this gives $l_c\sim 3-4$). 
When $H$ maps to $P_{\epsilon}$ in just a few
iterates, so that $l<l_c$, 
a significant difference is observed between 
$\mu_C(P_{\epsilon})$ and $\mu_N(P_{\epsilon})$ 
(this explains the wells in Fig. \ref{c_xsin}).

Coming back to the average lifetimes, 
if the hole does not map back 
to itself in just a few iterates, i.e., if the shortest 
periodic orbit within $H$ has a period larger than $l_c$, 
then $\mu_C(H)\simeq\mu_N(H)$, or simply $\tau\simeq\mu_N(H)^{-1}$. 
This explains the overall 
behavior of lifetimes (see Fig. \ref{brs_ill}). 
On the other hand, if $H$ encompasses a short periodic orbit 
(period$\equiv l(H)<l_c$), the two measures differ within the hole. 
Quantitatively, we substitute $P_{\epsilon}\rightarrow H$ in 
Eq. (\ref{recur}) and approximate  $e^{l(H)/\tau}\simeq 1$ and 
$\mu_C(M^{-l(H)})\simeq\mu_N(M^{-l(H)})$. 
This results in 

\begin{equation}
\tau\simeq (1-i(H))^{-1}\mu_N(H)^{-1}.   \label{str}
\end{equation}
Although $l(H)$ is small, the approximation 
$a(M^{-l(H)})\simeq 1$ is justified 
if $l(M^{-l(H)})>l_c$, or simply, if the period of the 
second shortest orbit within $H$ exceeds $l_c$. 
We have tested relation (\ref{str}) and consequently the 
approximations that lead to it in a number of systems. 
We have compared $\tau(H)$ with $\mu_N(H)^{-1}$ by 
changing the position of $H$ 
from "the most exceptional" point, 
the fixed point, to longer cycles. 
In Fig. \ref{valid} we display a test of Eq. (\ref{str}) for the 
the generalized baker's map (see Ref. \cite{O_b}, p. 75, 
$\lambda_a=0.35,\lambda_b=0.40,\alpha=0.40,\beta=0.60$),
and for the H\' enon map (see Ref. \cite{Hen} $a=1.4,b=0.3$). 
Recalling that $i(H)$ decreases exponentially with $l(H)$, and 
considering Eq. (\ref{str}), we see that the parameter 
$S(H)=(1-i(H))^{-1}$ decreases rapidly towards $1$ with 
the increase of $l(H)$ (see Fig. \ref{valid}). 
Eq. (\ref{str}) is robust and can be 
applied for holes of different shapes, 
as long as $\tau>>1$. 
If (for the 2D maps) we tailor the hole as a rectangle with sides 
of length $\epsilon$ parallel to the stable and unstable 
manifold segments, and center it on a short periodic orbit, 
then 

\begin{equation}
\tau\simeq(1-\Lambda_u^{-1})^{-1}\mu_N(H)^{-1}. \label{str1}
\end{equation}
$\Lambda_u$ denotes the magnitude of the expanding 
eigenvalue of that orbit. 
Eq. (\ref{str1}) also applies to 1D maps. 
Note that the approximation $i(H)\simeq\Lambda_u^{-1}$ 
assumes that the natural measure is smooth along 
the unstable direction within $H$. 
We observe that $\tau/\mu_N(H)^{-1}$ is independent 
of $\epsilon$. 
This is in accordance with the 
statement that the lifetime $\tau$ scales with $\epsilon$ 
just like $\mu_N(H)^{-1}$ \cite{OGY_1,Shin_1}.

Let us consider an application of Eq. (\ref{str1}). 
Suppose that we wish to control a chaotic system around an 
unstable fixed point. 
In order to obtain the position of the fixed point, 
its unstable eigenvalue, and other information required for 
the control, an observation of the free running system is 
needed\cite{OGY_1,DRS}. 
From this observation, we can also evaluate the visitation 
frequency to the $\epsilon$-vicinity of the fixed 
point, i.e. $\mu_N(H\equiv H_{\epsilon}(\vec\xi))$. 
$\epsilon$ is determined by the maximally allowed 
deviation of the control parameter from its nominal 
value\cite{OGY_1,DRS}. 
The question of interest is how many iterates 
are needed on the average ($\tau$), before a chaotic trajectory 
enters the region $H$, when the control becomes 
attainable\cite{OGY_1}. 
The prediction given by $\mu_N(H)^{-1}$ is an underestimate, 
since we are on the fixed point. 
For example, if the underlying dynamics of the 
system is the asymmetric tent map (with the same parameters as 
in Fig. \ref{brs_ill}), and 
if $H=(0.62111801\ldots-0.002,0.62111801\ldots+0.002)$, 
the estimate for the lifetime $\mu_N(H)^{-1}$ gives $250$ 
iterates. 
On the other hand, the numerically calculated lifetime is 
$\simeq 627$ iterates, which is more than twice as 
long. 
The lifetime obtained from formula (\ref{str1})
is $641$ iterates, which is very close to 
the numerically calculated lifetime. 
Thus, Eq. (\ref{str1}) can be utilized to 
easily and accurately obtain $\tau$ from 
an observation of the free running system.

In summary, we have studied the average lifetime ($\tau$) it takes 
for a randomly started orbit to land in a small region ($H$)
on a chaotic attractor. 
That problem was introduced in Ref. \cite{OGY_1} as 
an important issue for controlling chaos. 
Our main result is that if a low-period unstable periodic orbit 
visits the region $H$, then 
the lifetime $\tau$ significantly deviates 
from the inverse of the natural measure contained within $H$ 
($\mu_N(H)^{-1}$). 
The parameter of this deviation, $\tau/\mu_N(H)^{-1}$, 
is a function of the expanding eigenvalue of that low-period 
orbit.

\begin{figure}
\caption{$\tau(\xi)$ (solid line) and 
$\mu_N(H)^{-1}$ (dashed line) vs. $\xi$  for
a) the asymmetric tent map $(k_1^{-1}=0.39,\ k_2^{-1}=0.61)$, 
and b) the $\sin{\pi x}$ map. 
$H=(\xi-\epsilon,\xi+\epsilon),\ \epsilon=0.005$.}
\label{brs_ill} 
\end{figure}

\begin{figure}
\caption{Figure displays $\mu_C$ (solid line) in 
comparison to $\mu_N$ (dashed line) for 
the $\sin\pi x$ map ($\epsilon=0.005$).
Fig. 2a) shows $\mu_C$ for the hole positioned on 
$\xi=0.66$. 
The hole is not visited by a short periodic orbit. 
Note that $\mu_C(H)\simeq\mu_N(H)$. 
Fig. 2b) shows $\mu_C$ for the hole located on the 
fixed point. 
Note that $\mu_C(H)<\mu_N(H)$. 
In both figures $\mu_C$ and $\mu_N$ are globally identical, 
except at the first 3-4 images of the hole, which are plotted 
underneath the graphs.}
\label{c_xsin} 
\end{figure}

\begin{figure}
\caption{Numerically evaluated parameter $\tau/\mu_N(H)^{-1}$ 
for the Henon map (diamonds) and the Baker map (circles) in 
comparison to $(1-i(H))^{-1}$ (horizontal bars). 
The hole of radius $\epsilon=0.005$ is centered on the shortest periodic 
orbit (period$=l(H)$) visiting $H$.}
\label{valid} 
\end{figure}

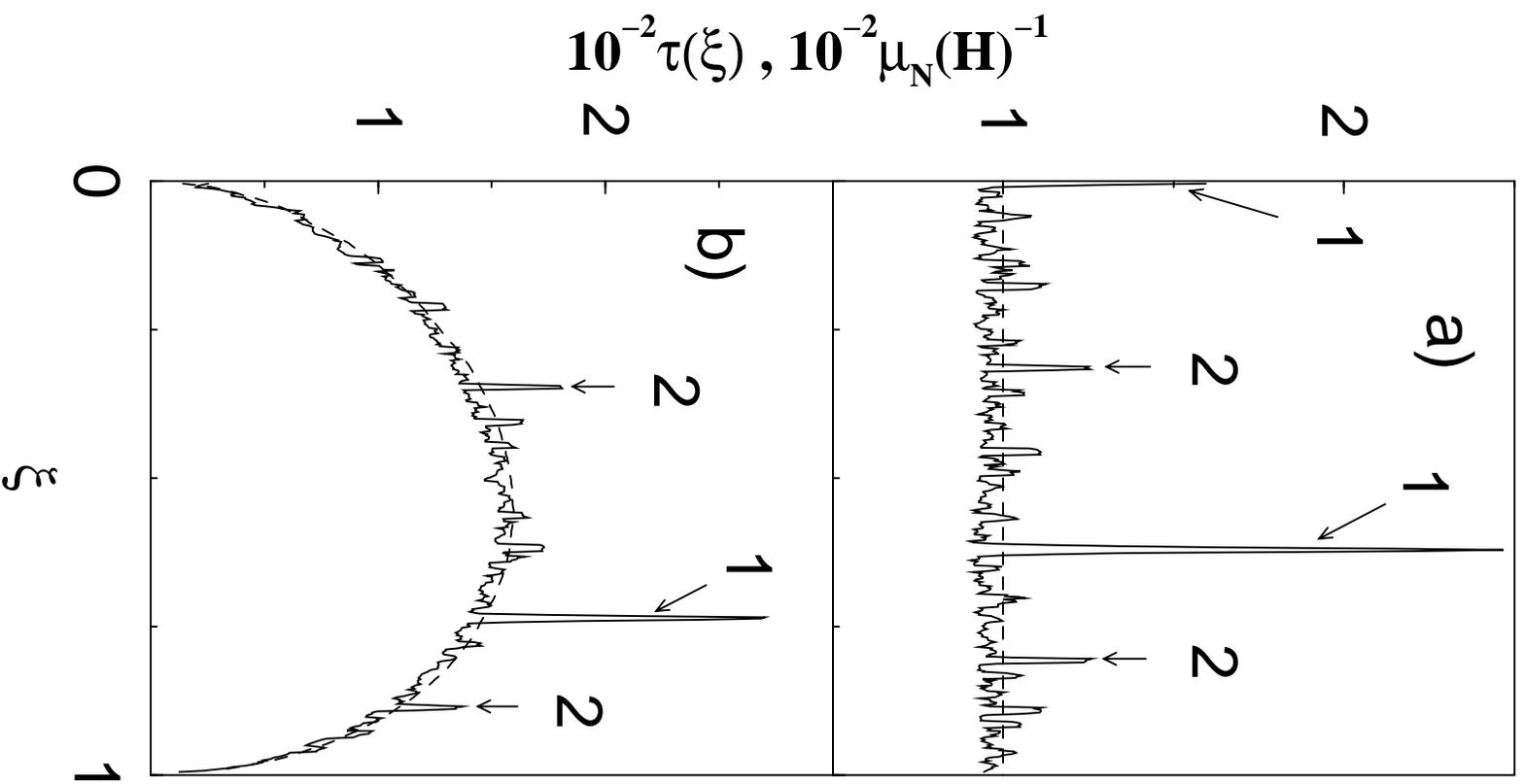
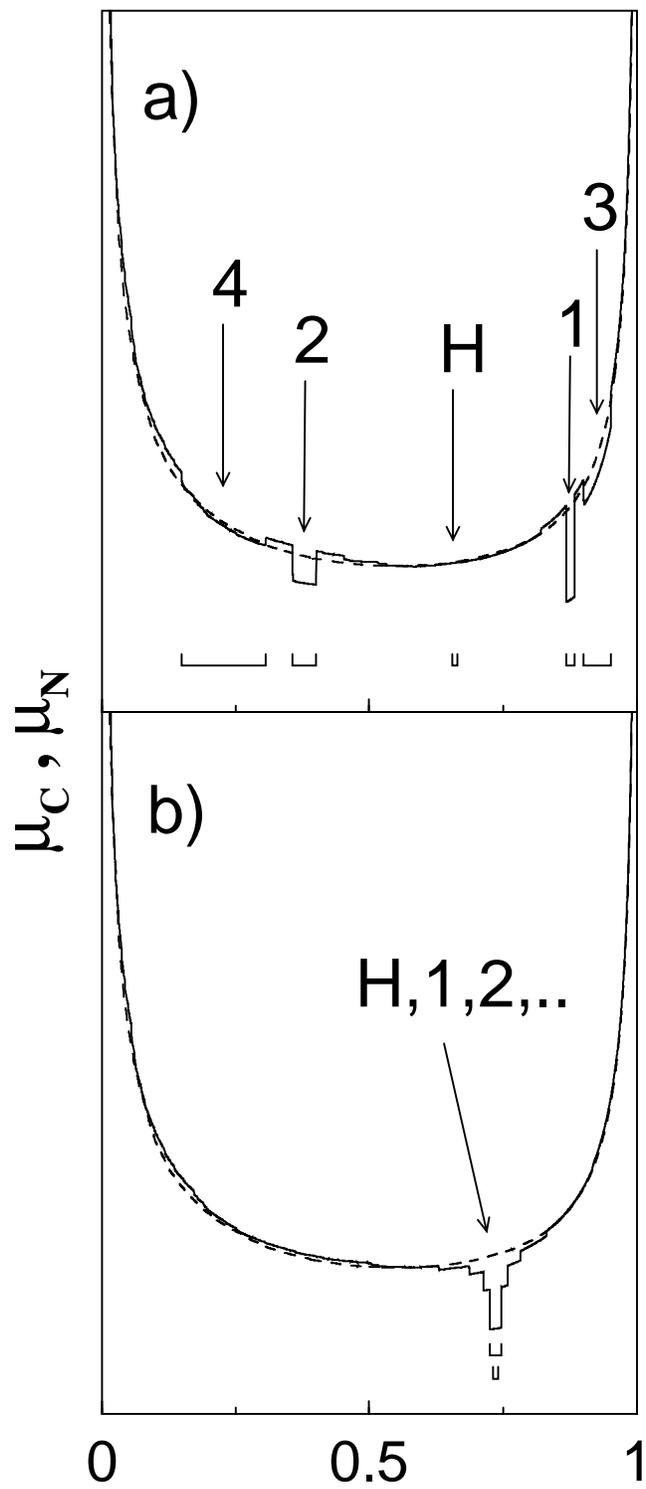
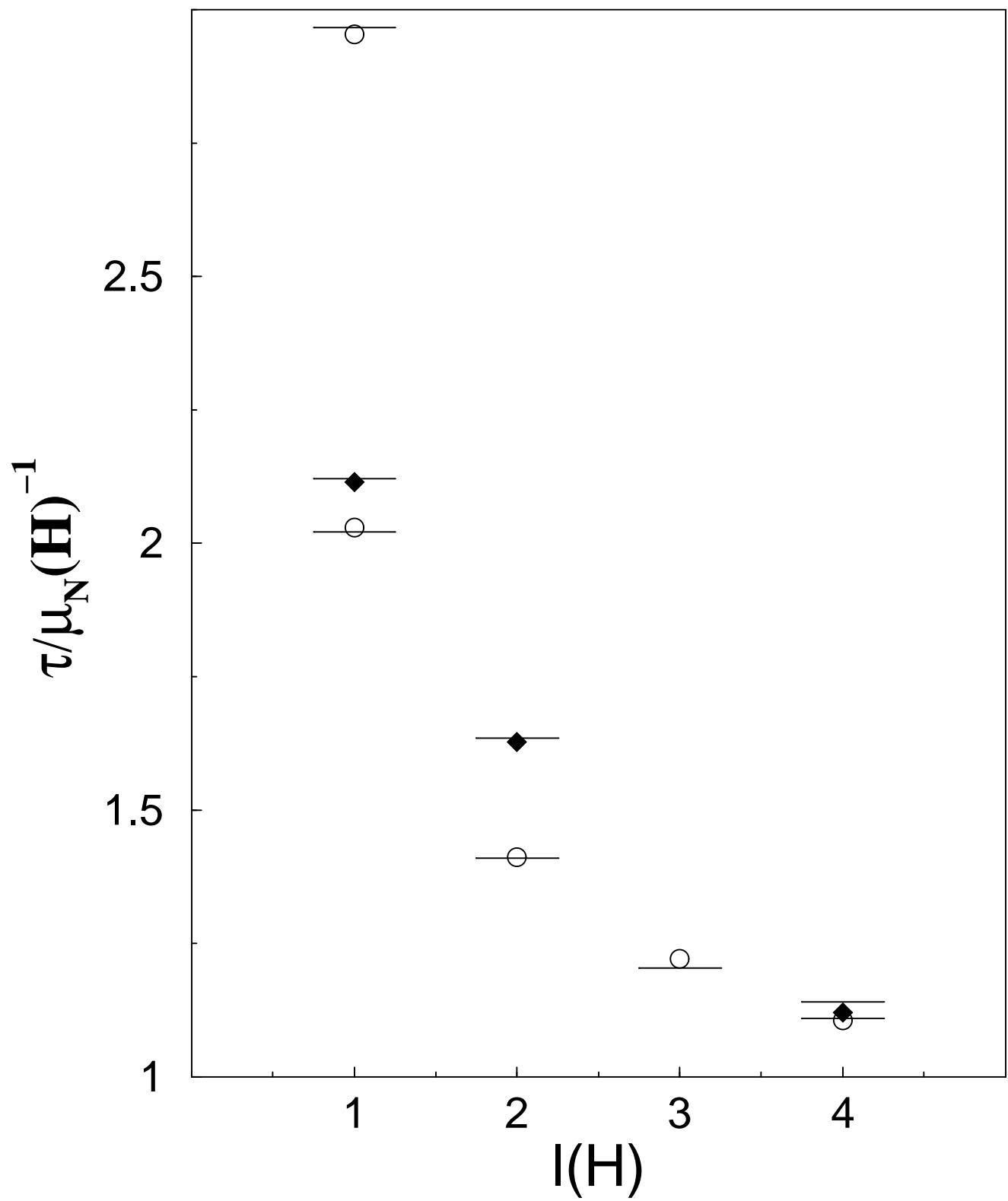
\end{document}